\documentclass[doublecol,figures,linenumbers]{epl2}

\usepackage{dcolumn}
\usepackage{bm}
\usepackage{color}
\usepackage{amsmath}
\usepackage{amssymb}

\title{Scaling of the Splash Threshold for Low-Viscosity Fluids}

\author{Cacey S. Stevens}
\institute{The James Franck Institute and Department of Physics, The University of Chicago, Chicago, Illinois 60637, USA}

\pacs{47.20.Ma}{Interfacial Instabilities}
\pacs{47.55 D-}{Drops and Bubbles}

\abstract{
The ambient gas pressure is determined for the onset of splashing of low-viscosity liquid drops on smooth dry surfaces as we change the control parameters: drop impact velocity, drop radius, viscosity, surface tension, density, and gas molecular weight. 
This threshold pressure indicates that there are two distinct regimes when drop impact velocity is varied.
By rescaling data using functions of only three dimensionless numbers, the commonly used Reynolds and Weber numbers, as well as the ratio of drop radius to gas mean free path,
 all data is collapsed to a single curve that encompasses both regimes.}

\begin{document}

\maketitle

\section{Introduction}
After years of study, physicists and engineers are still presented with a task of great difficulty: to provide criteria for the outcomes of drop impact in terms of all possible control parameters.
For the case of a liquid drop impacting a smooth dry surface, the drop may bounce, spread on the surface, or splash, emitting many smaller droplets \cite{Worthington, Tran}.
The velocity of impact and drop size, as well as liquid properties (viscosity, density, and surface tension) and surface roughness, have long been known to influence the outcome \cite{Yarin}.
Investigators have proposed conditions for the onset of splashing based solely on these parameters \cite{Stow, Mundo, Range, VanderWal}.
These are commonly expressed as functions of dimensionless numbers so that the criteria can lead to an understanding of the underlying mechanisms. 
However, there is little agreement among the proposed criteria, and they often contradict one another \cite{Rein}.

Only a few experimental studies of the splash threshold include the ambient gas pressure, $P$, which surprisingly is a crucial parameter for creating a splash \cite{Xu, Xu2, Liu, Ratner}.
Once $P$ is below a threshold value, a drop no longer splashes but spreads smoothly on the surface. 
Accordingly, splash criteria should include $P$ and gas molecular weight.	
Threshold pressure values reveal distinct regimes that occur at different values of surface roughness, liquid viscosity $\mu_L$, and impact velocity $u_0$ \cite{Xu, Xu2, Driscoll, Latka2}.
Therefore, splash criteria need to be determined for each regime separately.

Notably, Xu \textit{et al.} \cite{Xu} developed a splash criterion that described drop impact in the regime occurring at low-$\mu_L$ and high-$u_0$.
However, the measurements reported here expose a discrepancy in that scaling collapse when surface tension is varied;
thus a different scaling of the splash threshold for low-$\mu_L$ drop impact is presented.
All splash threshold data, at both low and high $u_0$, collapse cleanly onto a single curve when rescaled by three dimensionless numbers. 
The gas mean free path is important when describing the role of ambient gas in the collapse.
This master curve suggests a crossover between the low-$u_0$ and high-$u_0$ behavior and reveals different splash criteria for these two regimes.

\section{Experimental details} 
Experiments were conducted with ethanol, fluorinert, water-glycerol mixtures, and silicone oils with viscosities $\mu_L$ ranging from 0.5 mPa$\thinspace$s to 2.7 mPa$\thinspace$s, densities $\rho_L$ ranging from 750 kg/m$^3$ to 1860 kg/m$^3$, and surface tensions $\sigma$ ranging from 16 mN/m to 67 mN/m. 
I filmed drops of radius $R$ from $0.8 \pm  0.05$ mm to $2.0  \pm  0.1$ mm using a high speed camera (Phantom V12) as drops were released inside a transparent chamber from 0.15 m to 1.5 m above dry smooth glass slides (Fisher brand cover glass).
As shown previously \cite{Stow, Ratner}, the drop shape upon impact affects the subsequent splash formation. 
For example, drops that were oblate just before impact were more likely to splash. 
Therefore, only the impact of drops with a distortion in aspect ratio of less than 5\% from the spherical shape were considered in the results reported here.

The average surface roughness of the glass substrate is about 5 nm, measured using an atomic force microscope \cite{Latka}. 
A new glass slide was used for each drop impact to avoid contamination from residual liquid of previous splashes.
Side views recorded with the high-speed camera were used to determine the radius $R$ and impact velocity $u_0$.
Each measurement was repeated 4-8 times to confirm reproducibility.
$P$ was varied from 5 kPa to 101 kPa with one of three gases in the chamber: Helium (He), air, or Sulfur hexafluoride (SF$_6$). 
The gases have viscosities $\mu_G$ from 15.6 $\mu$Pa$\thinspace$s (SF$_6$) to 19.8 $\mu$Pa$\thinspace$s (He) and molecular weights $m_G$ from 4 Daltons (He) to 146 Daltons (SF$_6$) \cite{Weast}; 
$\mu_G$ does not vary with $P$ over the range studied.

\section{Splash threshold collapse} 
As a drop impacts a smooth surface, it ejects a liquid sheet which then breaks up into droplets (see left column of Fig. 1(a)).
No droplets emerge after impact if $P$ is below a threshold value; in that case, the drop simply spreads on the surface.
The boundary between splashing and spreading, shown in the right column of Fig 1(a), is defined in terms of a threshold pressure $P_T$, the pressure at which we first see the sheet break up into droplets.
The inset of Fig.\ 1(b) shows $P_T$ versus $u_0$ for 1.7 mPa$\thinspace$s silicone oil drops in an atmosphere of air ({\color{red}$\blacktriangle$}) and in an atmosphere of $SF_6$ ($\triangledown$). 
As with previous experiments \cite{Xu}, there are two regimes in $u_0$.
Initially $P_T$ decreases rapidly with $u_0$ to a minimum value at $u^*_0$.
Above $u_0^*$, $P_T$ increases then gradually decreases with $u_0$.
$P_T$ measurements in $SF_6$ are only shifted to lower pressures.
Data in both $u_0$-regimes fall onto a single curve when plotted in terms of a scaled pressure, $P_T(m_G/m_{air})^{0.5}$, 
as shown in the main plot of Fig.\ 1(b).
This is the same scaling as was found by Xu \textit{et al.} \cite{Xu}.

\begin{figure}[h] 
\onefigure[width=3 in]{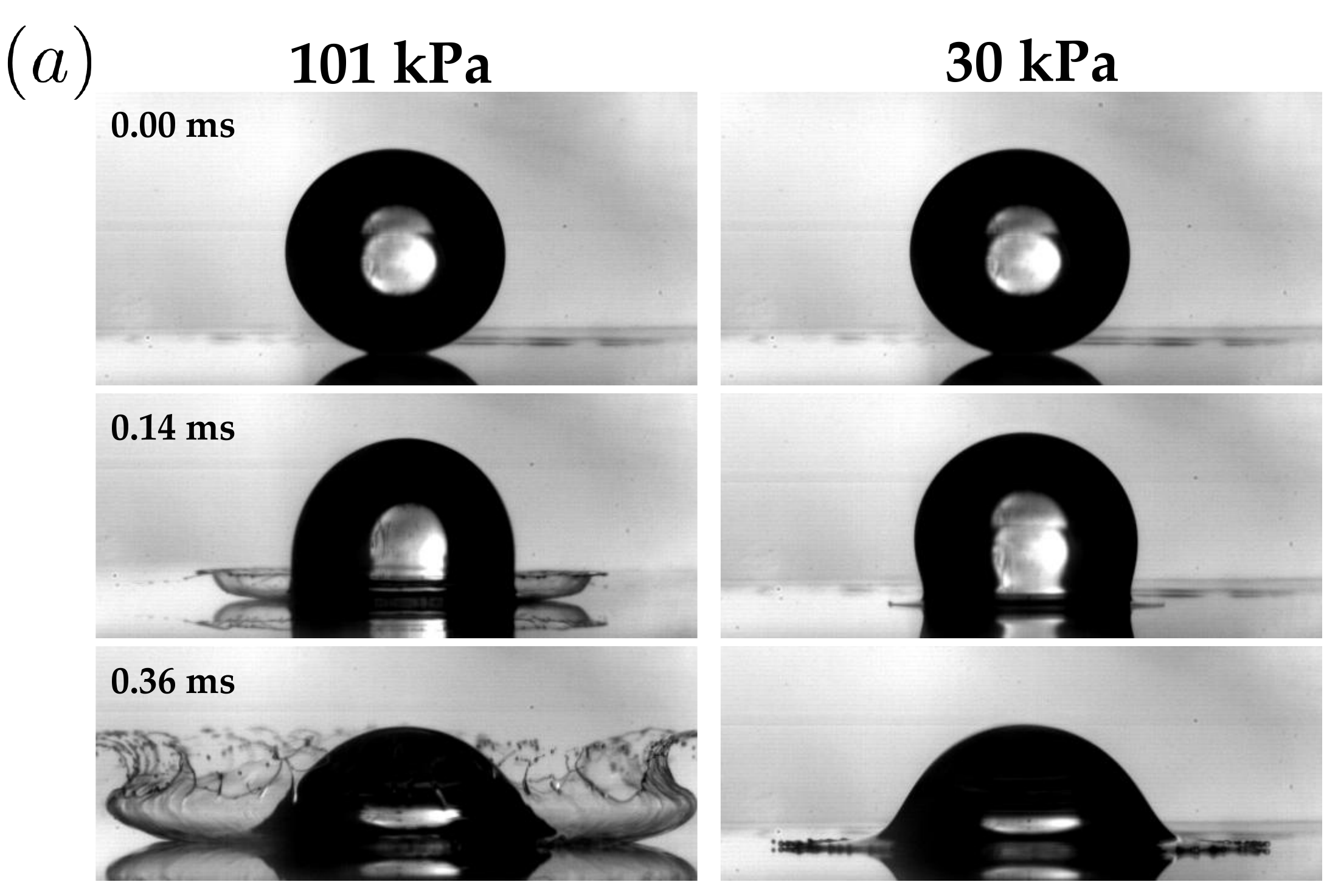}
\label{f1a}
\onefigure[width=3.1 in]{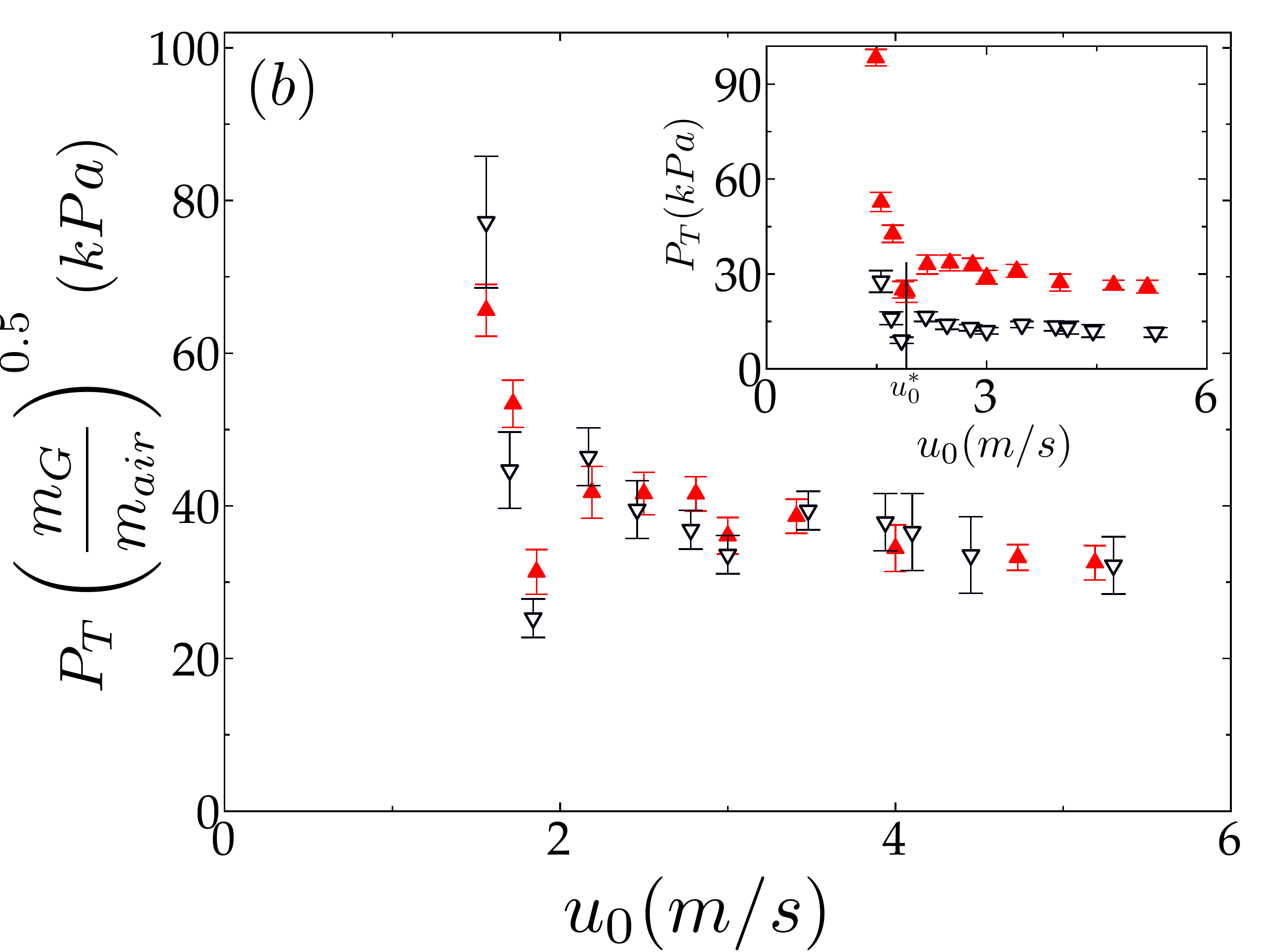}
\caption{(a) A 1.7 mPa$\thinspace$s silicone oil drop is shown impacting smooth glass at 101 kPa (left) and near $P_T$ at 30 kPa (right). Splashing is suppressed at low pressures. (b) \textit{Inset}: $P_T$ vs. $u_0$ of silicone oil drops (R = $1.6\pm0.1$ mm, $\mu_L$ = 1.7 mPa$\thinspace$s) in an atmosphere of air ({\color{red}$\blacktriangle$}) or $SF_6$ ($\triangledown$). The error bars indicate the pressure range for which the ejected sheet first breaks up into droplets. $u^*_0$ indicates the transition between low-$u_0$ and high-$u_0$ regimes. Higher $m_G$ ($SF_6$) lowers the curve but does not affect the trend in data. \textit{Main}: Scaled threshold pressure, $P_T(m_G/m_{air})^{0.5}$, versus $u_0$ for the two gases, collapsing data in both regimes.}
\end{figure}

In Fig.\ 2(a), $P_T$ versus $u_0$ is shown as control parameters are systematically changed: $R$, $\mu_L$, $\sigma$, $\rho_L$, and $m_G$.
All data sets display the same qualitative shape; curves simply move along the $P_T$ and $u_0$ axes.

\begin{figure*}[t] 
\centering
\includegraphics[width=15cm,height=10cm]{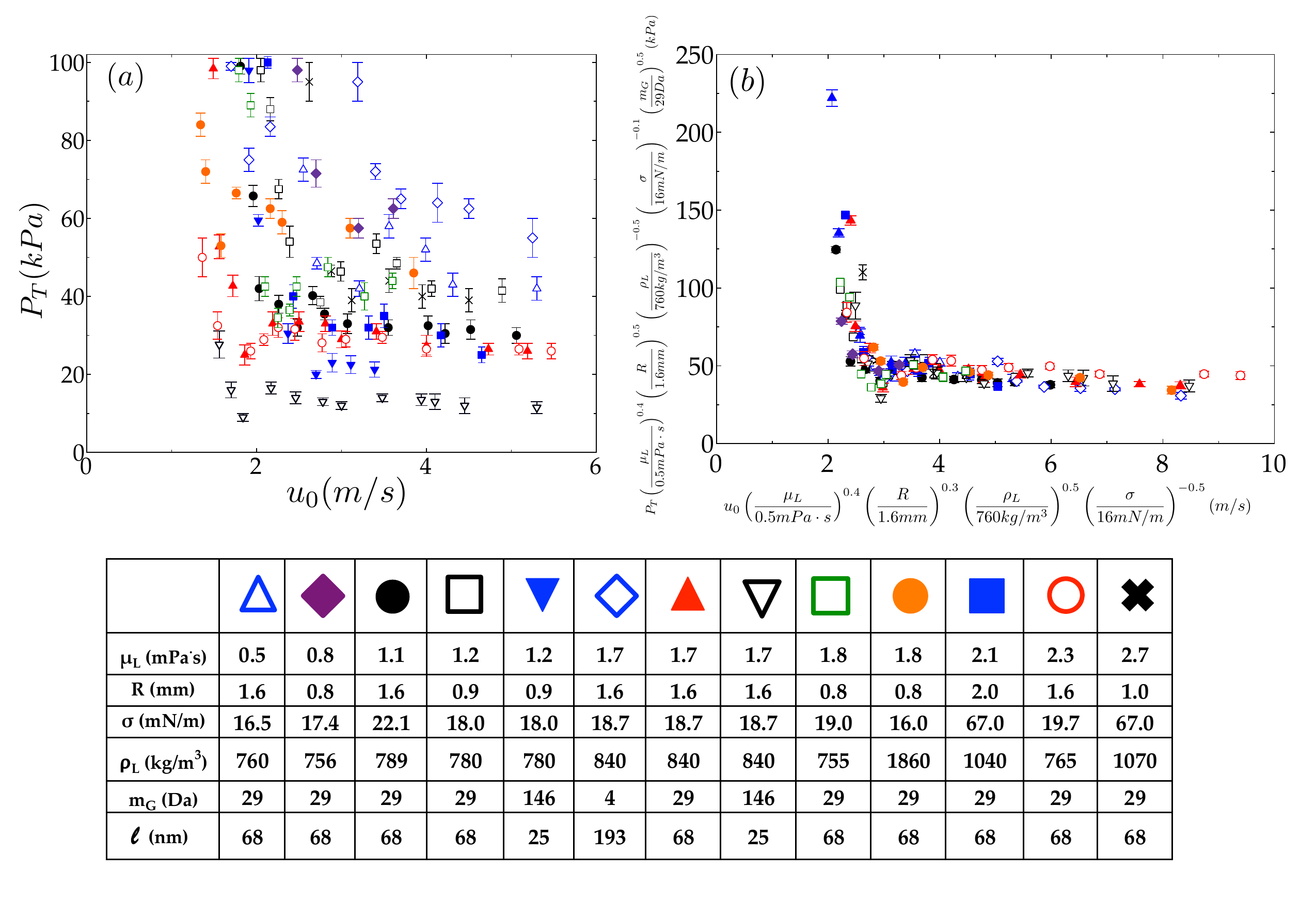}
\caption{(a) $P_T$ vs. $u_0$, varying $\mu_L$, $R$, $\sigma$, $\rho_L$, and $m_G$. Data sets correspond to symbols given in the table. (b) The data in both regimes collapse to one curve when $P$ and $u_0$ axes are scaled as functions of $\mu_L$, $R$, $\sigma$, $\rho_L$, and $m_G$ as given in Equations 1 and 2 of the text.}
\label{f2}
\end{figure*}

All data in both regimes of $u_0$ fall onto a single curve when the $u_0$ and $P_T$ axes are rescaled by functions of $R$, $\mu_L$, $\sigma$, $\rho_L$, and $m_G$ (see Fig. 2(b)): 

\begin{equation}
\label{e1}
u_{scaled} = u_0 \mu_L^{0.4} R^{0.3} \rho_L ^{0.5} \sigma^{-0.5} 
\end{equation}

\begin{equation}
\label{e2}
P_{scaled} = P_T \mu_L^{0.4} R^{0.5} \rho_L^{-0.5} \sigma^{-0.1} m_G^{0.5}
\end{equation} 

Each parameter was treated separately, leading to the scaled $u_0$ and $P_T$ axes.
It is challenging to find a rescaling of axes as a unique expression of all control parameters.
The experimental range of each parameter is limited, and there are slight fluctuations in the data.
Though not unique, these expressions lead to a good collapse of all data, over the obtainable range of the parameters.

It is insightful to find a scaling in terms of dimensionless numbers.
The liquid properties, impact velocity, and drop radius can be expressed as dimensionless numbers: the Reynolds number ($\rm{Re}$=$\rho_LRu_0/\mu_L$) giving the ratio of inertial to viscous forces and the Weber number ($\rm{We}$=$\rho_L R u_0^2/\sigma$) giving the ratio of the inertial to surface tension forces.
These experiments covered the range of $580<\rm{Re}<13100$ and $100<\rm{We}<2140$.

Another dimensionless parameter is introduced to encompass gas properties: the ratio of drop radius to the gas mean free path, $R/\ell$. 
Note that $P$ and $m_G$ are expressed through $\ell$ as $\ell^{-1}\propto Pm_G^{0.5}$ \cite{Weast}.
Since $\ell$ is inversely proportional to $P$, $\ell$ at the splash threshold is determined with the following relation: 

\begin{equation}
\label{e3}
\frac{\ell_T}{\ell_{atm}} = \frac{P_{atm}}{P_T}
\end{equation}

\noindent
where $\ell_T$ is the mean free path at $P_T$ and $\ell_{atm}$ is the mean free path of $He$, air, or $SF_6$ at atmospheric pressure ($P_{atm}$=101 $kPa$) and room temperature (values found in \cite{Weast, Marsh}).

\begin{figure}[h] 
\onefigure[width=2.9 in]{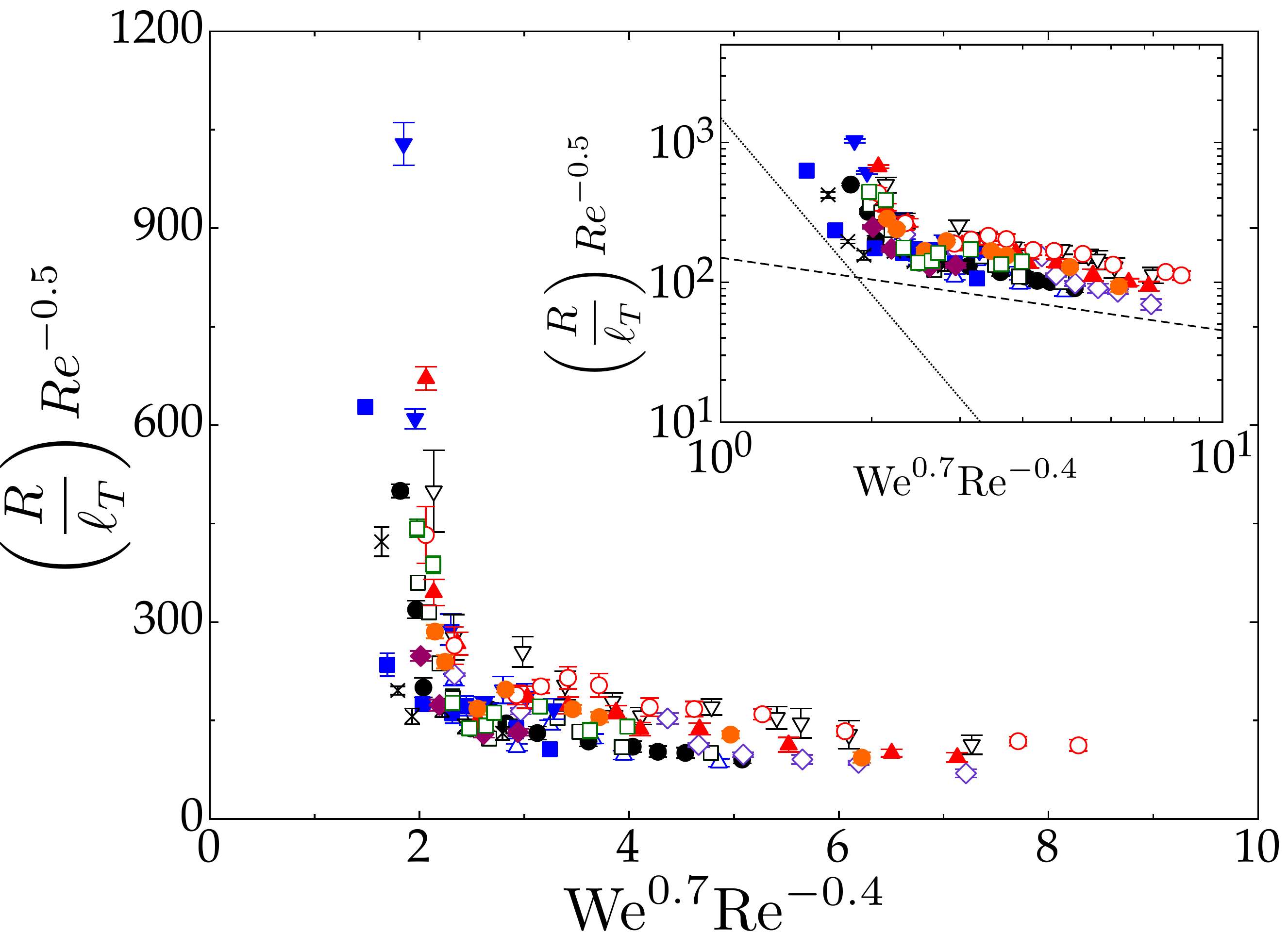}
\caption{Data are collapsed to a master curve by rescaling axes to functions of dimensionless numbers Re, We, and  $R/\ell_T$. Low-$u_0$ and high-$u_0$ regimes follow approximate power laws of $-4.2\pm0.85$ and $-0.52\pm0.09$, respectively, as shown in the inset plot of the data on a log-log scale.}
\label{f3}
\end{figure}

Using $R/\ell_T$, Re, and We, I replot all data of Fig. 2(a) using axes that are rescaled to a dimensionless pressure and velocity (see Fig. \ref{f3}): 

\begin{equation}
\label{e4}
P_T \rightarrow \left(\frac{R}{\ell_T}\right)^{1.0\pm0.2}\rm{Re}^{-0.5\pm0.08}
\end{equation}

\begin{equation}
\label{e5}
u_0 \rightarrow \rm{We}^{0.7\pm0.05} \rm{Re}^{-0.4\pm0.04}
\end{equation}
The errors indicate the range over which this decent collapse is achieved.
From the scaling of Fig. 2(b), the nearest collapse is found with dimensionless numbers, which encompass all physical quantities of the problem. I note that the collapse using dimensional variables (Fig. 2(b)) is better than that achieved with dimensionless quantities (Fig. 3). One clear reason for this can be attributed to the number of free parameters used in each case. In the collapse using dimensional numbers, six parameters are used whereas for the collapse with dimensionless variables, only half of the parameters are used; the relative simplicity of the collapse with dimensionless numbers is an achievement. It is possible to add additional dimensionless variables in order to improve the collapse. Although I have tried to include other parameters, the collapse achieved did not improve significantly.

\begin{figure}[h] 
\centering 
\onefigure[width=2.8 in]{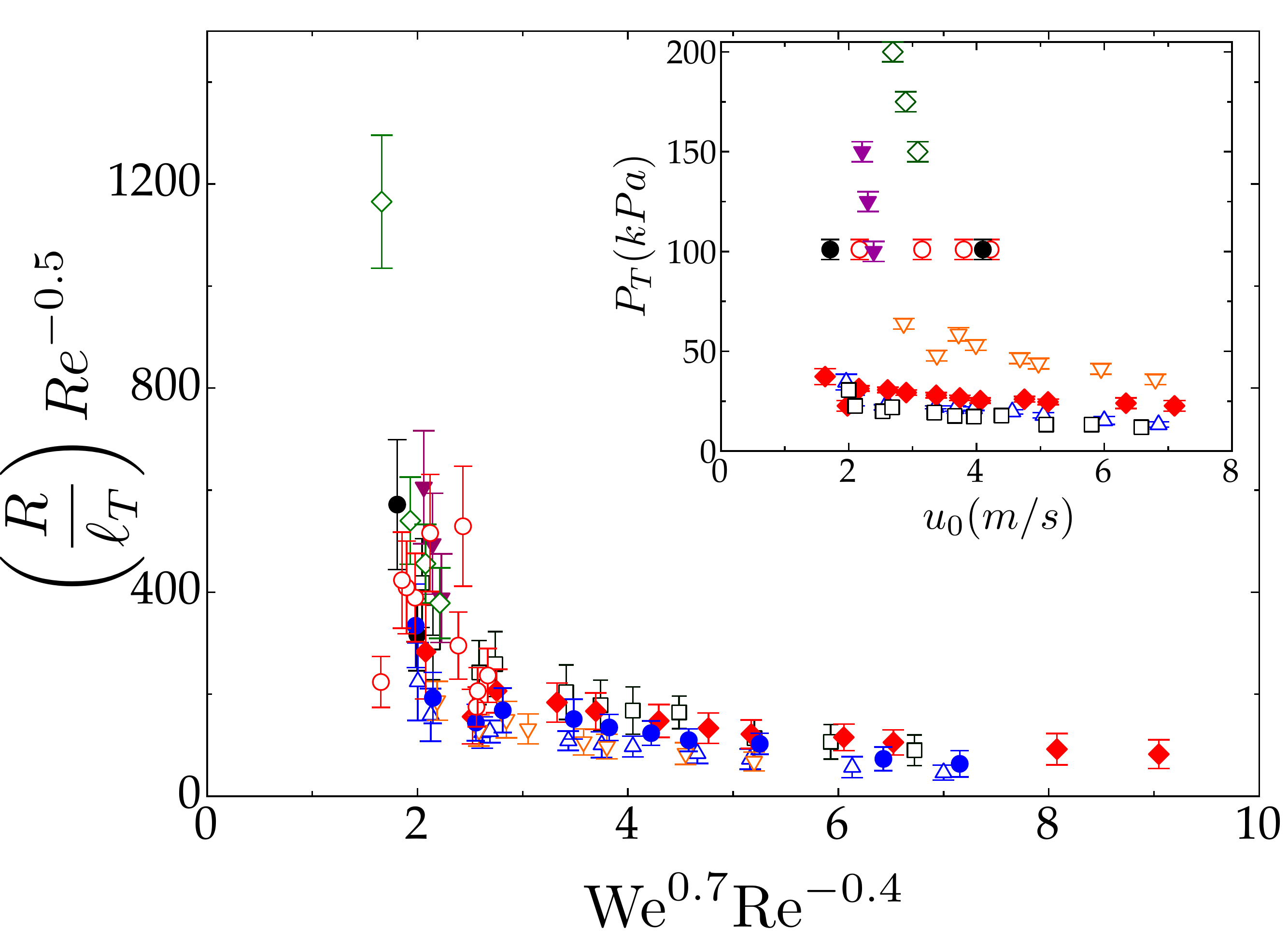}
\onefigure[width=3.1 in]{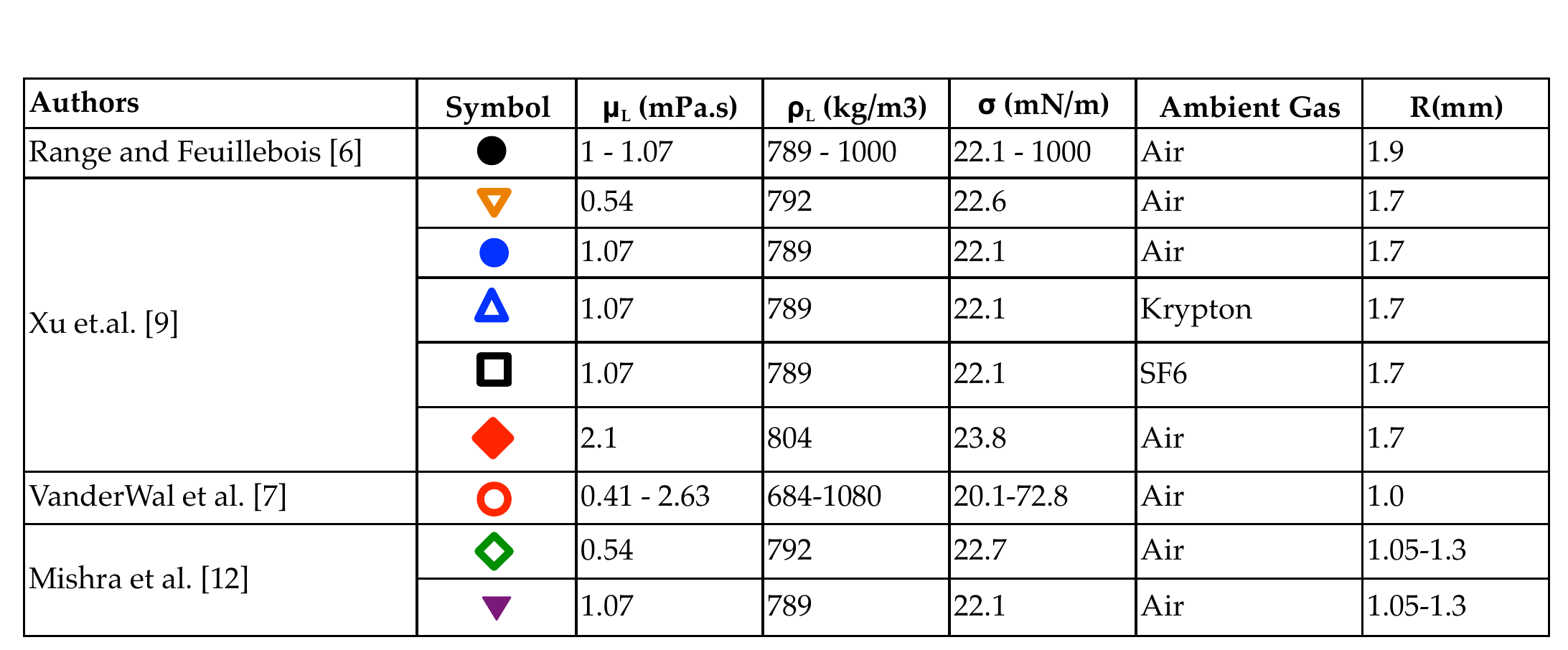}
\caption{\textit{Inset}: $P_T(u_0)$ for published splash threshold data of other authors. Data sets correspond to symbols noted in table. \textit{Main}: Data are scaled onto the master curve of the low-$\mu_L$ splash threshold. }
\label{f4}
\end{figure}

Low-$\mu_L$ splash threshold measurements of other authors are shown in the inset plot of Fig. \ref{f4}.
The scaling applies whether the splash threshold is determined by varying solely $\rm{Re}$ and $\rm{We}$ at $P_{atm} $\cite{Range, VanderWal}, decreasing $P$ \cite{Xu}, or increasing $P$ above $P_{atm}$ \cite{Ratner}.
All data collapse to the master curve, as shown in the main plot of Fig. \ref{f4}.

\section{Splash correlations}
Figure \ref{f3} clearly highlights two distinct regimes in $u_0$ with the crossover occurring at $u_0^*$.
From this curve, a splash criterion is found for each regime.
First consider the low-$u_0$ regime, for which there is no previous scaling attempt.
In this regime, the master curve approximately follows a power law of $-4.2$.
The data therefore scales as $(R/\ell_T)\cdot $$\rm{Re^{-0.5}}$ $\propto$ $\rm{(We^{0.7}Re^{-0.4})^{-4.2}}$
and can be simplified to the following relation:

\begin{equation}
\label{e6}
\left(\frac{R}{\ell_T}\right)^{0.5\pm0.1}\rm{Re^{-1.1\pm0.2}We^{1.5\pm0.3}}\approx 100
\end{equation}

In Fig. 5(a), Eq. \ref{e6} at threshold pressure is presented as a function of $P$ for the low-$u_0$ regime.
The errors in exponents are determined such that the average deviation of all data from the mean value is less than 25\%.
This correlation indicates that splashing is expected for values greater than the threshold.

The analysis is extended to the high-$u_0$ regime, where the curve approximately follows a power law of $-0.52$.
The data scales as $(R/\ell_T)\cdot \rm{Re^{-0.5}}$ $\propto$ $\rm{(We^{0.7}Re^{-0.4})^{-0.52}}$ and
 leads to the correlation at high-$u_0$:
 
\begin{equation}
\label{e7}
\left(\frac{R}{\ell_T}\right)^{0.5\pm0.1}\rm{Re^{-0.35\pm0.05}We^{0.2\pm0.03}}\approx20
\end{equation}

\noindent
The errors are determined such that the deviation from the mean is within 10\%.
The threshold value for the splash transition in the high-$u_0$ regime is shown as a function of $u_0$ in Fig 5(b).

This new threshold is not the same as the previously proposed model by Xu \textit{et al.} \cite{Xu}.
They derived an estimate based on variations to some, but not all, parameters.
After varying all liquid control parameters, including $\sigma$, $R$, and $\rho_L$, a new threshold is found for this regime.
For instance, a weaker effect of surface tension $\sigma$ ($P \propto \sigma^{0.4}$ compared to $P \propto \sigma$) is found,
though both criteria qualitatively show that higher $\sigma$ inhibits splashing.
There is also a weaker correlation with $u_0$; the dependence of $P$ on $u_0$ is lower than was found by Xu $\textit{et al.}$ \cite{Xu}. 
The proposed criterion does show the same dependence on $m_G$ and gas temperature as expressed through $\ell_T$.

\begin{figure}[h]
\centering
\includegraphics[width=2.8 in]{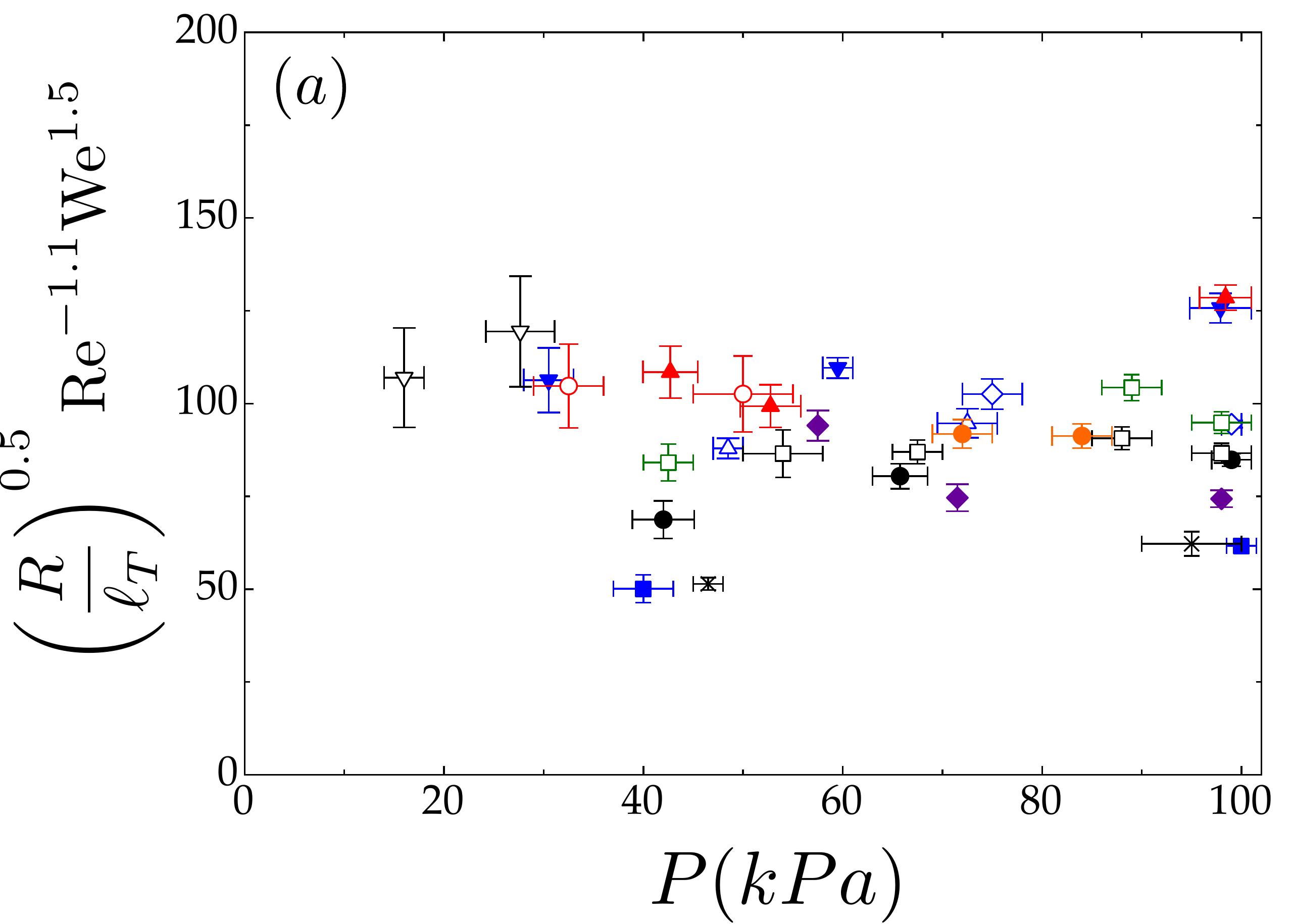}
\includegraphics[width=2.8 in]{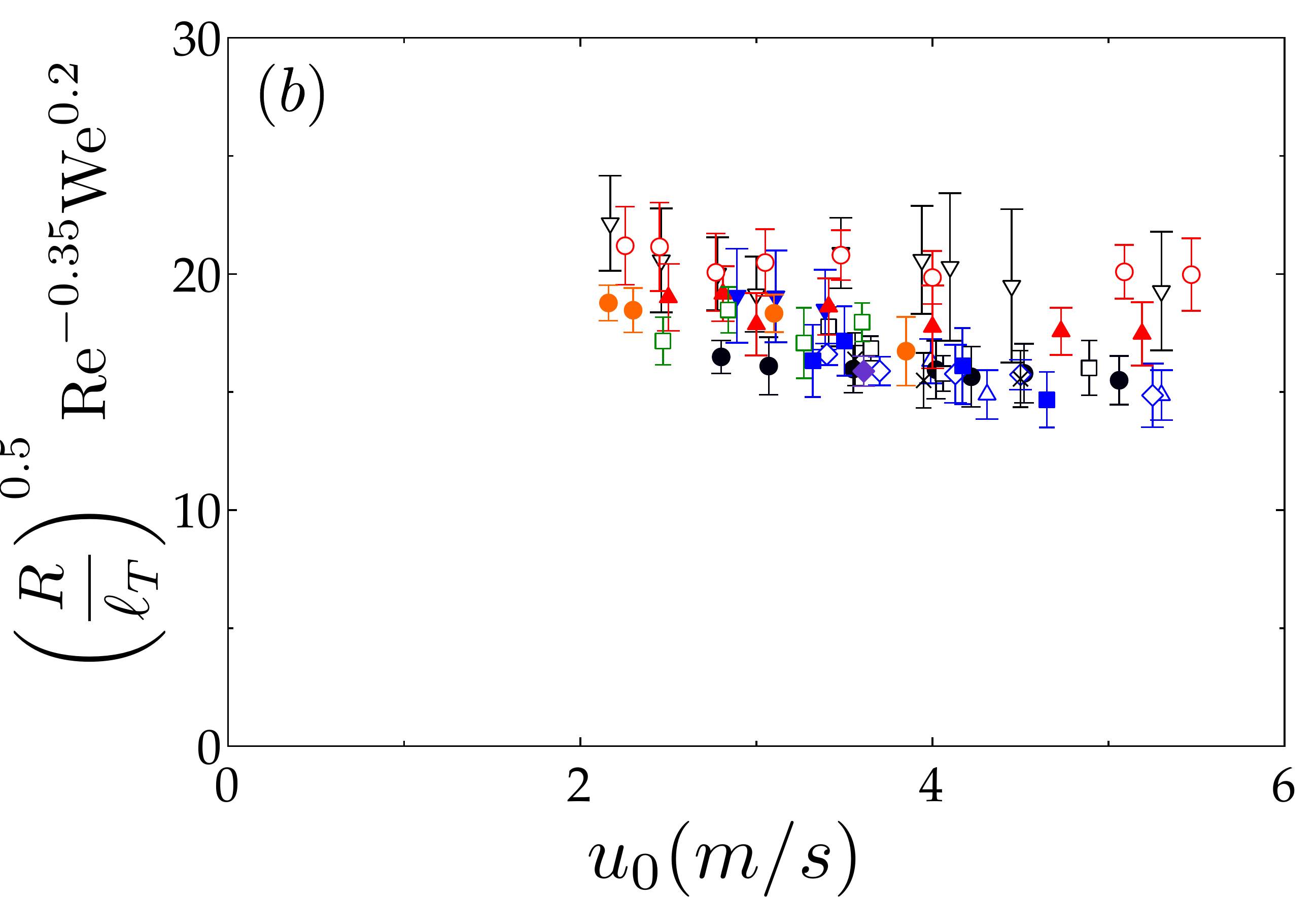}
\caption{Splash threshold criteria for (a) low-$u_0$ and (b) high-$u_0$ regimes. Data sets correspond to those of Fig. \ref{f2} and \ref{f3}.}
\end{figure}

\section{Conclusions}
I provide a comprehensive study of low-$\mu_L$ drop impact on dry, smooth surfaces.
The threshold pressure for splashing highlights two regimes in impact velocity $u_0$ with a crossover between them at $u_0^*$.
Remarkably all data, over both regimes in $u_0$,  can be rescaled onto a single curve, allowing us to define splash criteria approximately with only three dimensionless numbers: $R/\ell_T$, $Re$, and $We$.
A criterion is provided for the threshold in the low-$u_0$ regime, for which no data collapse was previously reported.
A new scaling for high-$u_0$ is found.

All work reported in this paper focuses on the low-$\mu_L$ regime.
Prior experiments describe a low-$\mu_L$ splash as the formation of an expanding sheet, which lifts up from the surface to a crown shape and becomes smaller at lower pressures \cite{Xu}.
This splash creation was proposed to result from air trapped under the drop upon impact \cite{Mani,Mandre}.
In the high-$\mu_L$ regime, a splash was shown to develop through thin sheet ejection from a thicker lamella; this thin sheet is ejected at later times with lower pressures \cite{Driscoll, Driscoll2}.
In addition to the low-$\mu_L$ splash thresholds presented here, it is valuable to find similar criteria for the velocity regimes at high viscosity.
Such scaling estimates for the regimes in liquid viscosity can lead to theory of the basic mechanisms behind these different splashes.

\acknowledgements
I am particularly grateful to Sidney R. Nagel for his keen insight and valuable guidance throughout this work.
I thank I. Bischofberger,  J. Burton, T. Caswell, D. Devendran, M. Driscoll, A. Latka, I. Peters, and W. Zhang for helpful discussions. 
This work was supported by NSF Grant DMR-1105145 and facilities of the University of Chicago NSF-MRSEC, supported by DMR-0820054. 
I also acknowledge support from the NSF Graduate Research Fellowship Program.

\end{document}